\documentclass[aps,prd,12point,twocolumn,nofootinbib,showpacs,superscriptaddress]{revtex4-2}

\usepackage{graphicx}
\usepackage{dcolumn}
\usepackage{tabu}
\usepackage{bbold}
\usepackage{bbm}
\usepackage{dsfont}
\usepackage{amsmath}
\usepackage{amssymb}
\usepackage{float}
\usepackage{natbib}
\usepackage{balance}
\usepackage{braket}
\usepackage{mathrsfs}
\usepackage{bm}
\usepackage{lipsum}
\usepackage{comment}
\usepackage{xfrac}
\usepackage[dvipsnames]{xcolor}
\usepackage[title]{appendix}
\usepackage{txfonts}
\usepackage{xcolor}
\usepackage{slashed}
\usepackage[normalem]{ulem}
\usepackage[breaklinks=true,pdfborder={0 0 0},bookmarksopen]{hyperref}
\hypersetup{
	colorlinks   = true, 
	urlcolor     = magenta, 
	linkcolor    = blue, 
	citecolor   = Red 
}

\begin{document}

\title{Neutrino flavor oscillations inside matter in conformal coupling models}

\author{Fay\c{c}al Hammad}
\email{fhammad@ubishops.ca}
\affiliation{Department of Physics \& Astronomy,
Bishop's University,\\
2600 College Street, Sherbrooke, QC, J1M 1Z7, Canada}
\affiliation{Physics Department, Champlain College-Lennoxville,\\
2580 College Street, Sherbrooke, QC J1M 0C8, Canada}

\author{Parvaneh Sadeghi}
\email{psadeghi20@ubishops.ca}
\affiliation{Department of Physics \& Astronomy,
Bishop's University,\\
2600 College Street, Sherbrooke, QC, J1M 1Z7, Canada}

\author{Nicolas Fleury}
\email{nfleury22@ubishops.ca}
\affiliation{Department of Physics \& Astronomy,
Bishop's University,\\
2600 College Street, Sherbrooke, QC, J1M 1Z7, Canada}

\begin{abstract}
We recently studied neutrinos flavor oscillations in vacuum within conformal coupling models. In this paper, we extend that analysis by investigating neutrino flavor oscillations inside matter within a general conformal coupling scenario. We first derive the general formula for the flavor transition probability inside matter in arbitrary static and spherically symmetric spacetimes. The modified resonance formula of the MSW effect is derived and the corresponding adiabaticity parameter of the effect is extracted. An application of our results to the case of two-flavor neutrinos within the well-known chameleon and symmetron conformal coupling models is made.
\end{abstract}

\maketitle
\section{Introduction}\label{sec:Intro}
Among the stringent requirements in any modified gravity model that attempts to solve the dark matter and/or dark energy problems \cite{DarkReview} is to be also able to reproduce every solar-system measurement conducted so far. Indeed, as modified gravity models usually introduce extra invariant curvature terms, extra geometric degrees of freedom and extra scalar fields into the Einstein-Hilbert action, deviations from Newtonian physics in those models could easily sneak in. In order to avoid the appearance in those models of any unwanted ``fifth force'' interaction, the latter has to either be non-existent or simply become too small to be detected at the solar-system scale. The latter possibility is known as screening. Among the very well studied models that offer a screening mechanism are the chameleon model \cite{Chameleon} and the symmetron model \cite{Symmetron} (see also Ref.\,\cite{ChameleonTests} for a more recent review and discussion of the various feasible experimental tests on such a screening mechanism). 

These two models, which belong to the so-called scalar-tensor family of modified gravity theories, contain a scalar field $\phi(x)$ that couples to matter fields $\Psi_i(x)$ through the spacetime metric $g_{\mu\nu}$. More specifically, one replaces in those models any curved-spacetime matter Lagrangian $\mathcal{L}_m(\Psi_i,g_{\mu\nu})$ by the Lagrangian $\mathcal{L}_m(\Psi_i,\tilde{g}_{\mu\nu})$, where the metric $\tilde{g}_{\mu\nu}$ is related to the spacetime metric $g_{\mu\nu}$ of the gravitational sector by a Weyl conformal transformation \cite{Wald}: $\tilde{g}_{\mu\nu}=\Omega^2(\phi)g_{\mu\nu}$. The nonvanishing conformal factor $\Omega(\phi)$ is a functional of the scalar field $\phi(x)$. The conformal factor has the form $\Omega(\phi)=\exp(\beta\phi)$ in the chameleon model \cite{Chameleon}, whereas in the symmetron model it has the form $\Omega(\phi)=1+\beta\phi^2$ \cite{Symmetron}. The arbitrary constant $\beta$ has the dimensions of an inverse mass and inverse mass-squared, respectively. 

The fifth force appears within these models as an extra term in the geodesic equation of massive bodies, and it is proportional to the derivative ${\rm d}[\ln\Omega(\phi)]/{\rm d}\phi$. The screening mechanism in the chameleon model is realized thanks to the increasing mass of the scalar field with the environment's density, which leads to an extremely short-range interaction. In the symmetron model, the screening mechanism is realized thanks to the variation of the scalar field's coupling to matter, which becomes very weak within dense environments. It is therefore very important to study within these models other well-known physical phenomena ---\,such as those that are at the center of multi-messenger astrophysics \cite{MultiReview1,MultiReview2}\,--- that do not rely directly on the fifth force appearing in the geodesic equation. Our aim in this paper is to examine the fate of neutrino flavor oscillations inside matter and in the presence of a strong gravitational field under the conformal coupling scenario.

Since the very first theoretical proposal for neutrino flavor transition \cite{Pontecorvo1,Pontecorvo2} and its experimental discovery soon after \cite{HistoryNO0, HistoryNO1,HistoryNO2,HistoryNO3,HistoryNO4,HistoryNO5,HistoryNO6,HistoryNO7,HistoryNO8,HistoryNO9,HistoryNO10,HistoryNO11}, neutrinos have taken a special place in modern cosmology and astrophysics \cite{PhenomenologyWithN}. In fact, as neutrinos turned out to have non-vanishing masses which are responsible for giving rise to the observed neutrino flavor oscillations, studying the latter has gradually become indissociable from cosmology and gravitational physics \cite{PRL2004,PRL2005,NPB2005,JHP,JCAP2004,PRD2003,PLB2007,PRD2017,JCAP2018,DarkU2020,PRD2021a,PRD2021b,DarkU2021,Arxiv2022}.

Although neutrino flavor oscillation is caused by the mere propagation of neutrinos, either in vacuum or inside matter, the matter effect at resonance considerably amplifies the flavor transition probability, and it has been the subject of intense experimental investigation. Neutrino flavor oscillation is induced inside matter via the forward scattering of neutrinos with the fermions of the medium through which they propagate \cite{NBook}. This forward scattering manifests itself as the well-known resonance in the Mikheyev-Smirnov-Wolfenstein (MSW) effect \cite{MSW1,MSW2}. Therefore, if neutrinos are subjected to a conformal coupling the fermions with which such neutrinos interact should also be subjected to a conformal coupling. In addition, as the scalar field's profile in these screening models depends on the environment's density, we expect a more interesting effect of the scalar field on the flavor oscillations inside matter. Investigating the matter effect on flavor oscillations of conformally coupled neutrinos becomes thus of equal importance, but of higher priority than that of investigating the effect of gravity on the flavor oscillations in vacuum. We shall therefore conduct a rigorous study of the MSW effect within the general framework of the conformal coupling scenario before applying it to the specific chameleon and symmetron cases.

The remainder of this paper is organized as follows. In Sec.\,\ref{Sec:Vacuum}, we briefly recall the derivation of the flavor transition probability for conformally coupled neutrinos propagating in vacuum under the influence of a gravitational field described by a general static and spherically symmetric metric. The main tools needed later for dealing with the propagation of neutrinos in curved spacetime are introduced in this section. In Sec.\,\ref{Sec:Matter}, we generalize that formula for the transition probability by including the effect of the interaction of those neutrinos with matter. In Sec.\,\ref{Sec:Twoflavor}, we apply to the case of two-flavor neutrinos the general formulas obtained in Sec.\,\ref{Sec:Matter}. We consider throughout this paper mainly the case of a universal conformal coupling, but we discuss in Sec.\,\ref{Sec:AppB} the way the results of Sec.\,\ref{Sec:Matter} generalize further to the case of a non-universal conformal coupling. We summarize our main findings in the brief section \ref{Sec:Conclusion}.
\section{Flavor oscillations in vacuum}\label{Sec:Vacuum}
A neutrino flavor $\ket{\nu_\alpha}$ is made out of a superposition of three mass eigenstates $\ket{\nu_j}$ via the Pontecorvo-Maki-Nakagawa-Sakata (PMNS) unitary mixing matrix $U^*_{\alpha j}$ \cite{NBook}: $\ket{\nu_\alpha}=\sum_jU_{\alpha j}^*\ket{\nu_j}$. On the other hand, the curved-spacetime Dirac equation describing the dynamics of a spinor wavefunction $\psi_j$ associated to a neutrino mass eigenstate of mass $m_j$ reads $(i\gamma^\mu\nabla_\mu-m_j)\,\psi_j=0$\footnote{We shall work with the spacetime metric signature $(-,+,+,+)$, and we shall set throughout the paper $\hbar=c=1$.}. This equation is obtained from the usual Dirac equation by (see, e.g., Ref.\,\cite{QFT2}): replacing the flat-space gamma matrices $\gamma^a$ by the curved-spacetime gamma matrices $\gamma^\mu=e^\mu_a\gamma^a$ and by replacing the partial derivatives $\partial_\mu$ by the spin covariant derivatives $\nabla_\mu=\partial_\mu+\frac{1}{8}\,\omega_\mu^{\,ab}[\gamma_a,\gamma_b]$\footnote{We use the Latin letters $(a,b)$ to denote flat-spacetime indices while we reserve the Greek letters $(\mu,\nu)$ for curved-spacetime indices. Mass eigenstates will be denoted using the letters $(j,k)$ and neutrino flavors will be denoted by the Greek letters $(\alpha,\beta)$.}. The spin connection $\omega_\mu^{\,ab}$ is related to the Christoffel symbols $\Gamma_{\mu\nu}^\lambda$ and the vierbeins $e^a_\mu$ by $\omega_\mu^{\,\,ab}=e^a_\nu\partial_\mu e^{\nu b}+\Gamma_{\mu\nu}^\lambda e^{a}_\lambda e^{\nu b}$. The curved-spacetime Dirac equation is obtained from the following fermion-matter Lagrangian:
\begin{equation}
\mathcal{L}_m=\int{\rm d}^4x\sqrt{-g}\,\left[\tfrac{i}{2}\bar{\psi}_j\gamma^\mu\nabla_\mu\psi_j-\tfrac{i}{2}\nabla_\mu(\bar{\psi}_j)\gamma^\mu\psi_j-m_j\bar\psi_j\psi_j\right].
\end{equation}
Here, the metric determinant $\sqrt{-g}$ is that of the metric $g_{\mu\nu}$, and $\bar\psi$ is the adjoint spinor: $\bar\psi=\psi^\dagger\gamma^0$. 

For neutrinos conformally coupled to the metric via a scalar field $\phi$, their Lagrangian is built using the metric $\tilde{g}_{\mu\nu}=\Omega^2(\phi)g_{\mu\nu}$, the curved-spacetime gamma matrices $\tilde{\gamma}^\mu=\Omega(\phi)\gamma^\mu$ and the covariant derivatives $\tilde{\nabla}_\mu$ that are associated to the metric $\tilde{g}_{\mu\nu}$. We assume throughout the paper that $\phi$ and $g_{\mu\nu}$ are time independent. Since the fermion Lagrangian is conformally invariant \cite{QFT2}, for conformally coupled neutrinos the Lagrangian can be written in terms of the original spacetime metric $g_{\mu\nu}=\Omega^{-2}(\phi)\tilde{g}_{\mu\nu}$ as follows: 
\begin{align}\label{LagrangianDirac}
\!\!\!\!\!\mathcal{\tilde{L}}_m&=\!\int{\rm d}^4x\sqrt{-\tilde{g}}\left[\tfrac{i}{2}\bar{\psi}_j\tilde{\gamma}^\mu\tilde{\nabla}_\mu\psi_j-\tfrac{i}{2}\tilde{\nabla}_\mu(\bar{\psi}_j)\tilde{\gamma}^\mu\psi_j-m_j\bar\psi_j\psi_j\right]\nonumber\\
&=\int{\rm d}^4x\sqrt{-g}\!\left[\tfrac{i}{2}\tilde{\bar{\psi}}_j\gamma^\mu\nabla_\mu\tilde{\psi}_j-\tfrac{i}{2}\nabla_\mu(\tilde{\bar{\psi}}_j)\gamma^\mu\tilde{\psi}_j-\tilde{m}_j\tilde{\bar\psi}_j\tilde{\psi}_j\right].
\end{align}
The new effective mass $\tilde{m}_j$ and the new spinor field $\tilde{\psi}_j$ in this Lagrangian are given in terms of the original constant mass $m_j$ and the original spinor field $\psi_j$ by, respectively \cite{QFT2,SchroDi},
\begin{equation}\label{ConformalMassPsi}
    \tilde{m}_j=\Omega(\phi)\,m_j, \qquad \tilde{\psi}_j(x)=\Omega^{\frac{3}{2}}(\phi)\,\psi_j(x).
\end{equation}
The mass-shell condition satisfied by the conformally coupled mass eigenstates within the metric $g_{\mu\nu}$ is $g_{\mu\nu}\tilde{p}_j^\mu\tilde{p}_j^\nu=-\tilde{m}_j^2$, where the four-momentum $\tilde{p}^\mu_j$ carries the effect of the scalar field $\phi$ through the relation $\tilde{p}_j^\mu=\Omega(\phi)p_j^\mu$.
The four-momentum $p_j^\mu=m_j{\rm d}x^\mu/{\rm d}\tau$ is what a neutrino mass eigenstate $\ket{\nu_j}$ of mass $m_j$ would have if it were not conformally coupled.

Therefore, a neutrino mass eigenstate $\ket{\nu_j}$ conformally coupled to spacetime and emitted at a point $(0,\vec{x}_A)$ can be described at a detection point $(t,\vec{x}_B)$ by the following plane wave\footnote{For the wavepacket approach to neutrino flavor oscillations in curved spacetime, see Refs.\,\cite{Submitted,NWPG} and references therein.} \cite{Submitted}:
\begin{equation}\label{PlaneWave}
\ket{\nu_j(t,\vec{x}_B)}=e^{i\,\Phi_j}\ket{\nu_j},
\end{equation}
where $\ket{\nu_j}$ are the orthonormal basis of the Hilbert space of mass eigenstates, satisfying $\braket{\nu_j|\nu_k}=\delta_{jk}$. The quantum phase $\Phi_j$ is evaluated using the Stodolsky prescription \cite{Stodolsky} for computing the accumulated quantum phases of propagating quantum particles in curved spacetimes \cite{Submitted}:
\begin{equation}\label{ConformalAccumultatedPhase}
\Phi_j=-\int_A^B\tilde{m}_j\,{\rm d}\tau=\int_A^Bg_{\mu\nu}\tilde{p}_j^\mu\,{\rm d}x^\nu.
\end{equation}
Therefore, the probability amplitude $\braket{\nu_\beta|\nu_\alpha(t,\vec{x}_B)}=\bra{\nu_\beta}U^*_{\alpha j}\ket{\nu_j(t,\vec{x}_B)}$ for a conformally coupled neutrino produced at point $(0,\vec{x}_A)$ as an $\alpha$ flavor to be detected at the point $(t,\vec{x}_B)$ as a $\beta$ flavor is simply, $\sum_j\,U^*_{\alpha j}U_{\beta j}\exp(i\,\Phi_{j})$.

We shall consider for definiteness and simplicity in the remainder of this paper a radial propagation of neutrinos within a static {\it and} spherically symmetric metric by choosing the following general form for the latter:
\begin{equation}\label{Metric}
{\rm d}s^2=-\mathcal{A}(r)\,{\rm d}t^2+\mathcal{B}(r)\,{\rm d}r^2+r^2\left({\rm d}\theta^2+\sin^2\theta\,{\rm d}\varphi^2\right).
\end{equation}
Here, the arbitrary functions $\mathcal{A}(r)$ and $\mathcal{B}(r)$ of the radial coordinate $r$ are everywhere regular except maybe at a singularity or on a horizon. We let these functions be arbitrary for the sake of generality.
Within this spacetime metric, the energy $\tilde{E}_j(\tilde{p})$ and momentum $\tilde{p}_j$ of a conformally coupled and radially propagating mass eigenstate of mass $\tilde{m}_j$, as perceived by an observer at infinity, are given by $\tilde{E}_j(\tilde{p})=-g_{00}\tilde{p}_j^0=\mathcal{A}(r)\,\tilde{m}_j{\rm d} t/{\rm d}\tau$
and $\tilde{p}_j=g_{rr}\tilde{p}_j^r=\mathcal{B}(r)\,\tilde{m}_j{\rm d}r/{\rm d}\tau$. Furthermore, the existence of the time-like Killing vector $K^\nu=(1,0,0,0)$ implies the conserved energies $E_j(p)$ of each mass eigenstate:
\begin{equation}\label{ConservedEnergy}
E_j(p)=-g_{\mu\nu}K^\mu m_j\frac{{\rm d}x^\nu}{{\rm d}\tau}=\frac{\tilde{E}_j(\tilde{p})}{\Omega(\phi)}.
\end{equation}
These conserved energies $E_j(p)$ are the energies an observer at infinity would perceive if the neutrino mass eigenstates were propagating in the spacetime of metric $g_{\mu\nu}$. Combining these identities with the mass-shell condition $-\tilde{m}_j^2=g_{\mu\nu}\tilde{p}_j^\mu \tilde{p}_j^\nu$, leads to
\begin{equation}\label{MassShell}
-\tilde{m}_j^2=-\frac{\tilde{E}_j^2(p)}{\mathcal{A}(r)}+\tilde{m}_j^2\mathcal{B}(r)\left(\frac{{\rm d}r}{{\rm d}\tau}\right)^2,
\end{equation}
from which we may extract the following relation between $\tilde{p}_j$ and $\tilde{E}_j(\tilde{p})$ \cite{Submitted}:
\begin{align}\label{pinE(p)}
\tilde{p}_j&=\tilde{E}_j(\tilde{p})\sqrt{\frac{\mathcal{B}(r)}{\mathcal{A}(r)}}\left(1-\frac{\tilde{m}_j^2\mathcal{A}(r)}{\tilde{E}^2_j(\tilde{p})}\right)^{\frac{1}{2}}\nonumber\\
&\approx\sqrt{\frac{\mathcal{B}(r)}{\mathcal{A}(r)}}\left(\tilde{E}_j(\tilde{p})-\frac{\tilde{m}_j^2\mathcal{A}(r)}{2\tilde{E}_j(\tilde{p})}\right).
\end{align} 
We expanded here the square root up to the first order in the ratio $\tilde{m}_j^2/\tilde{E}_j(\tilde{p})$. Plugging this expression into Eq.\,(\ref{ConformalAccumultatedPhase}), the accumulated phase of each mass eigenstate at the detection point $(t,\vec{x}_B)$ takes the following form:
\begin{align}\label{ComputedPlanePhase}
\Phi_{j}&=-E_j(p)\int_{A}^{B}\Omega(\phi)\,{\rm d}t+E_j(p)\int_{r_A}^{r_B}\Omega(\phi)\sqrt{\frac{\mathcal{B}(r)}{\mathcal{A}(r)}}\,{\rm d}r\nonumber\\
&\quad-\frac{m_j^2}{2E_j(p)}\int_{r_A}^{r_B}\Omega(\phi)\sqrt{\mathcal{A}(r)\mathcal{B}(r)}\,{\rm d}r.
\end{align}
Here, use has been made of identities (\ref{ConformalMassPsi}) and (\ref{ConservedEnergy}), which allowed us to move the conserved energy $E_j(p)$ and the position-independent mass $m_j$ out of the integrals. To compute the transition probability based on expression (\ref{ComputedPlanePhase}) of the quantum phase, one has two different options \cite{Submitted}. In the first, one may leave the time integral as an unknown parameter to be integrated over; leading to a time-averaged flavor transition probability. The second option consists in turning the time integral into a radial integral by expressing ${\rm d}t$ in terms of ${\rm d}r$. It was shown in Ref.\,\cite{Submitted} that while both approaches lead to the same final result for the transition probability, only the second option is adequate when treating neutrinos as wavepackets. As we shall see in Sec.\,\ref{Sec:Matter}, it turns out that when dealing with neutrinos interacting with matter, only the second option is also adequate.

Using that ${\rm d}t/{\rm d}\tau=\tilde{E}_j(\tilde{p})/[\mathcal{A}(r)\,\tilde{m}_j]$
and ${\rm d}r/{\rm d}\tau=\tilde{p}_j/[\mathcal{B}(r)\,\tilde{m}_j]$, as well as Eq.\,(\ref{pinE(p)}), we find the following link between the coordinate elements ${\rm d}t$ and ${\rm d}r$ along the path of the $j$-th mass eigenstate: 
\begin{equation}\label{dr/dt}
{\rm d}t=\sqrt{\frac{\mathcal{B}(r)}{\mathcal{A}(r)}}\left(1-\frac{\tilde{m}_j^2\mathcal{A}(r)}{\tilde{E}_j^2(\tilde{p})}\right)^{-\frac{1}{2}}{\rm d}r\approx\sqrt{\frac{\mathcal{B}(r)}{\mathcal{A}(r)}}\left(1+\frac{\tilde{m}_j^2\mathcal{A}(r)}{2\tilde{E}_j^2(\tilde{p})}\right){\rm d}r.
\end{equation}
In the second step we expanded again the square root up to the first order in $\tilde{m}_j^2/\tilde{E}_j^2(\tilde{p})$. Now, since the detection position is $r=r_B$ for all of the mass eigenstates, a common detection time may be obtained only if one turns all time integrals into spatial integrals using a {\it unique} link between the elements ${\rm d}t$ and ${\rm d}r$. As the wave front of the most massive mass eigenstate is the last one to arrive at $r=r_B$, the time integral that needs to be performed in Eq.\,(\ref{ComputedPlanePhase}) for any mass eigenstate should be the time integral corresponding to the {\it most} massive state of the three mass eigenstates. Therefore, by denoting the mass and energy of the most massive state by $\tilde{m}_*$ and $\tilde{E}_*$, respectively, the time integral that should be inserted into the calculation (\ref{ComputedPlanePhase}) of the phase for all the mass eigenstates is
\begin{align}\label{MassiveTime}
    \int_A^B\Omega(\phi){\rm d}t&=\int_{r_A}^{r_B}\Omega(\phi)\sqrt{\frac{\mathcal{B}(r)}{\mathcal{A}(r)}}{\rm d}r\nonumber\\
    &\quad+\frac{m_*^2}{2E_*^2(p)}\int_{r_A}^{r_B}\Omega(\phi)\sqrt{\mathcal{A}(r)\mathcal{B}(r)}{\rm d}r.
\end{align}
In the second term on the right-hand side we moved out of the integral the position-independent ratio $\frac{1}{2}\tilde{m}_*^2/\tilde{E}_*^2(\tilde{p})=\frac{1}{2}m_*^2/E_*^2(p)$. Inserting this result for the time integral into the second line of Eq.\,(\ref{ComputedPlanePhase}), the resulting quantum mechanical phase of the $j$-th mass eigenstate reads \cite{Submitted}:
\begin{equation}\label{trPhase}
    \Phi_j=-\left[\frac{E_j(p)m_*^2}{2E_*^2(p)}+\frac{m_j^2}{2E_j(p)}\right]\int_{r_A}^{r_B}\Omega(\phi)\sqrt{\mathcal{A}(r)\mathcal{B}(r)}\,{\rm d}r.
\end{equation}
With this expression, the probability amplitude for an $\alpha$ flavor emitted at the fixed radial coordinate $r_A$ to be detected at a certain radial coordinate $r_B$ as a $\beta$ flavor takes the form
\begin{multline}\label{VacuumTransitionAmplitudes}
    \braket{\nu_\beta|\nu_\alpha(r_B)}=\sum_j\,U^*_{\alpha j}U_{\beta j}\exp\Bigg[-i\left(\frac{m_*^2}{2E_0}+\frac{m_j^2}{2E_0}\right)\\
    \times\int_{r_A}^{r_B}\Omega(\phi)\sqrt{\mathcal{A}(r)\mathcal{B}(r)}\,{\rm d}r\Bigg].
\end{multline}
We have approximated here the conserved energy $E_j(p)$ of each mass eigenstate by the common average energy $E_0$ of massless neutrinos. In fact, if we went up to the first order in $m^2/E_0$ by writing $E_{j}(p)\approx E_0+\xi m_j^2/2E_0$, where $\xi$ is a dimensionless constant smaller than one \cite{NBook}, we would only have ended up in our expressions with correction terms that are of the second order in $m^2/E_0$. The dimensionless constant $\xi$ depends entirely on the characteristics of the interaction and on the nature of the other particles taking place in the production process of neutrinos \cite{Giunti}. 

In the next section we shall generalize this result to include the interaction of neutrinos with matter.

\section{Flavor oscillations inside matter}\label{Sec:Matter}
As a neutrino flavor propagates inside matter, it interacts with the fermions of the medium through coherent forward elastic charged currents (CC) and neutral currents (NC) scatterings. Neutrino flavor oscillations are rather affected by the CC scattering, which gives rise to an effective potential $V_{\rm CC}$ inside a homogeneous and isotropic gas of unpolarized electrons. Indeed, the effective potential $V_{\rm NC}$ arising from the NC scattering is factored out as a constant common phase to all the flavors, and hence does not contribute to neutrino flavor oscillations \cite{NBook}. 

Now, these interactions of neutrinos with matter are best described using a quantum field theoretical analysis. In fact, the effective position-dependent potential is given by $V_{\rm CC}(x)=2\sqrt{2}\,G_F\,n_e(x)$, where $G_F$ is Fermi's constant and $n_e(x)$ is the number density of the electrons of the medium. This effective potential arises from a low-energy effective charged-current Lagrangian density $\mathcal{L}_{\rm CC}^{\rm eff}$. This Lagrangian density, describing the CC scattering of an electron-flavor neutrino off an electron inside the medium, is given by the following expression \cite{NBook}:
\begin{multline}\label{EffLagrangian}
    \mathcal{L}_{\rm CC}^{ \rm eff}=\frac{G_F}{\sqrt{2}}\sqrt{-g}\,g_{\mu\nu} \left[\bar{\psi}_e(x)\gamma^\mu\left(1-\gamma^5\right)\psi_{\nu_e}(x)\right]\\
    \times\left[\bar{\psi}_{\nu_e}(x)\gamma^\nu\left(1-\gamma^5\right)\psi_e(x)\right].
\end{multline}
Note that in this section we shall work mainly within the flavor basis. The electron and the electron-neutrino fields are denoted here by $\psi_e$ and $\psi_{\nu_e}$, respectively. The matrix $\gamma^5=i\gamma^0\gamma^1\gamma^2\gamma^3$ is the constant flat-spacetime gamma matrix. One extracts the effective potential $V_{\rm CC}(x)$ from this Lagrangian density by first using the Fierz identity to rearrange the fermion fields and then averaging the electron's field bilinears over the volume of the medium \cite{NBook}.

Recall now that the procedure for dealing with conformally coupled matter consists of two steps. In the first step, one replaces everywhere inside a given matter Lagrangian the spacetime metric $g_{\mu\nu}$ and all the derived geometric terms in the Lagrangian by, respectively, the conformally rescaled metric $\tilde{g}_{\mu\nu}$ and the corresponding conformally transformed geometric terms. In the second step, one maps back the rescaled metric $\tilde{g}_{\mu\nu}$ and the derived geometric terms in the Lagrangian into the original forms they had within the metric $g_{\mu\nu}$, but rescales all the matter fields and their masses according to the prescription (\ref{ConformalMassPsi}). In order to deal with flavor oscillations of conformally coupled neutrinos, one would then start from the Lagrangian (\ref{EffLagrangian}) and replace everywhere in it the fermion fields by their conformally coupled versions. Therefore, when it comes to Fermi's constant, one is tempted to just assign to it a conformal weight that would make the Lagrangian (\ref{EffLagrangian}) conformally invariant like the Dirac Lagrangian. This is indeed the strategy adopted in Ref.\,\cite{Arxiv2022}.

However, we know that the Lagrangian (\ref{EffLagrangian}) is itself extracted from a scattering amplitude obtained from a tree-level diagram involving the $W$ bosons (see, e.g., Ref.\,\cite{STModel}). In fact, the Fermi constant $G_F$ is given in terms of the weak-coupling constant\footnote{We denote the weak coupling constant here by $g_W$ instead of the usual $g$ in order to distinguish it from the spacetime metric determinant $g$.} $g_W$ and the mass $m_W$ of the $W$ bosons by $G_F=\frac{\sqrt{2}}{8}g^2_W/m_W^2$. This expression, in turn, arises from approximating the $W$ boson propagator in Minkowski spacetime by $i\eta_{\mu\nu}/m_W^2$ at the low energies of interest here, which are much smaller than the mass $m_W$. On the other hand, based on the propagator of massive vector fields in curved spacetime  \cite{Barvinsky}, for weak curvature the $W$ boson propagator can be written as
\begin{align}\label{Propagator}
    G_{\mu\nu}^W(x-y)&=\braket{0|{\rm T}[W_\mu(x)W^\dagger_\nu(y)]|0}\nonumber\\
    &=-i\int\frac{{\rm d}^4p}{(2\pi)^4}\frac{g_{\mu\nu}+\frac{p_\mu p_\nu}{m^2_W}}{p^2+m_W^2-i\epsilon}e^{ip\,\cdot(x-y)}+\mathcal{O}(R),
\end{align}
where $\rm T$ stands for the time-ordered product and $\mathcal{O}(R)$ stands for neglected terms that are of the first order and higher in the spacetime curvature scalar and curvature tensors. Therefore, in order to deduce the new form the effective Lagrangian (\ref{EffLagrangian}) takes for conformally coupled matter, our strategy here will be to rather start from the more primitive tree-level interaction involving the propagator (\ref{Propagator}). 

Using the conformal metric $\tilde{g}_{\mu\nu}=\Omega^2(\phi)g_{\mu\nu}$, the $S$-matrix element for a conformally coupled left-handed electron-neutrino scattering from a conformally coupled electron of the medium can be written at tree level as follows,
\begin{multline}\label{ConfLagrangian}
i\frac{g^2_W}{8}\int\sqrt{-\tilde{g}}\,{\rm d}^4x\int\!\sqrt{-\tilde{g}}\,{\rm d}^4y \left[\bar{\psi}_e(x)\tilde{\gamma}^\mu\left(1-\gamma^5\right)\psi_{\nu_e}(x)\right]\\
\times\tilde{G}_{\mu\nu}^W(x-y)\left[\bar{\psi}_{\nu_e}(y)\tilde{\gamma}^\nu\left(1-\gamma^5\right)\psi_e(y)\right].
\end{multline}
We denoted here the propagator of the conformally coupled $W$ bosons by $\tilde{G}_{\mu\nu}^W(x-y)$. The latter is explicitly given by
\begin{align}\label{ConformalPropagator}
    \tilde{G}_{\mu\nu}^W(x-y)&=\braket{\tilde{0}|{\rm T}[W_\mu(x)W^\dagger_\nu(y)]|\tilde{0}}\nonumber\\
    &=-i\int\frac{{\rm d}^4p}{(2\pi)^4}\frac{\tilde{g}_{\mu\nu}+\frac{p_\mu p_\nu}{m^2_W}}{p^2+m_W^2-i\epsilon}e^{ip\,\cdot(x-y)}\nonumber\\
    &=\braket{0|{\rm T}[\tilde{W}_\mu(x)\tilde{W}^\dagger_\nu(y)]|0}\nonumber\\
    &=-i\int\frac{{\rm d}^4\tilde{p}}{(2\pi)^4}\frac{g_{\mu\nu}+\frac{\tilde{p}_\mu \tilde{p}_\nu}{\tilde{m}^2_W}}{\tilde{p}^2+\tilde{m}_W^2-i\epsilon}e^{i\tilde{p}\,\cdot(x-y)}\nonumber\\
    &\approx-\frac{ig_{\mu\nu}}{\tilde{m}_W^2}\delta^4(x-y).
\end{align}
In the first step, we introduced the conformal vacuum $\ket{\tilde{0}}$ corresponding to the metric $\tilde{g}_{\mu\nu}$ \cite{QFT2}, and in the second step we switched back to the metric $g_{\mu\nu}$ and took care of replacing the $W$-boson fields, their momenta and their mass by their conformal versions. This step is justified by the fact that the $W$-boson fields obey the massive Proca field equation which is conformally invariant when their mass $m_W$ transforms according to the first identity in Eq.\,(\ref{ConformalMassPsi}) \cite{What?,SchroDi}. In fact, a vacuum state for a conformally coupled field remains a vacuum state in the conformal spacetime \cite{QFT2}. In the last step, we used the low-energy limit approximation of the propagator. Plugging this approximation into the $S$-matrix element  (\ref{ConfLagrangian}), and then expressing everywhere in it the metric $\tilde{g}_{\mu\nu}$ in terms of $g_{\mu\nu}$ and the $\tilde{\gamma}^\mu$'s in terms of the $\gamma^\mu$'s, and using the second identity in Eq.\,(\ref{ConformalMassPsi}) for the fermion fields, the effective Lagrangian (\ref{EffLagrangian}) takes the following form for conformally coupled neutrinos:
\begin{multline}\label{ConfEffLagrangian}
    \mathcal{\tilde{L}}_{\rm CC}^{ \rm eff}=\frac{g_W^2}{8\tilde{m}_W^2}\sqrt{-g} \left[\bar{\tilde{\psi}}_e(x)\gamma_\mu\left(1-\gamma^5\right)\tilde{\psi}_{\nu_e}(x)\right]\\
    \times\left[\bar{\tilde{\psi}}_{\nu_e}(x)\gamma^\mu\left(1-\gamma^5\right)\tilde{\psi}_e(x)\right].
\end{multline}
We therefore see that the corresponding effective Fermi constant for conformally coupled neutrinos is $\tilde{G}_F=\frac{\sqrt{2}}{8}g_W^2/\tilde{m}_W^2=\Omega^{-2}(\phi)G_F$. When recalling that the physical dimensions of the Fermi constant are those of energy times volume (i.e., a mass dimension of $-2$), this result for the conformal transformation of the Fermi constant is indeed the naturally expected one according to the first identity in Eq.\,(\ref{ConformalMassPsi}).

Next, from the effective Lagrangian (\ref{ConfEffLagrangian}) we deduce that the effective position-dependent and $\phi$-dependent potential that needs to be used for conformally coupled neutrinos inside matter is
\begin{equation}\label{ConfVCC}
    \tilde{V}_{\rm CC}(x,\phi)=\sqrt{2}\,\tilde{G}_F\tilde{n}_e(x)=\sqrt{2}\,\Omega(\phi)G_Fn_e(x).
\end{equation}
In the second step we used the fact that the electron number density $\tilde{n}_e(x)$ of the conformally coupled background electrons within the metric $g_{\mu\nu}$ is related to the number density $n_e(x)$ within the metric $\tilde{g}_{\mu\nu}$ by $\tilde{n}_e(x)=\Omega^3(\phi)n_e(x)$. This identity stems from the definitions $n_e(x)={\rm d}N_e/{\rm d}\tilde{V}$ and $\tilde{n}_e(x)={\rm d}N_e/{\rm d}V$ of the number densities in the two frames for the unique total number of electrons ${\rm d}N_e$ in an infinitesimal spatial volume, and from the conformal transformation ${\rm d}\tilde{V}=\Omega^3(\phi){\rm d}V$ of any arbitrary infinitesimal spatial volume ${\rm d}V$. Although it is more natural to work with the density $\tilde{n}_e(x)$ than the density $n_e(x)$, we shall work with the latter rather than the former as it makes our formulas less cumbersome. Nevertheless, we shall express our important results (\ref{2FlavorElectronDensity}) and (\ref{ThetaMConstant}) in terms of $\tilde{n}_e(x)$ in order to explicitly see the effect of the conformal coupling. An explicit derivation of identity (\ref{ConfVCC}) for the effective potential $\tilde{V}_{\rm CC}(x)$ is given in Appendix \ref{Sec:AppA}. 

With this expression of the effective potential at hand, we proceed now to find the transition amplitude. Let us denote by a column vector $\Psi(r)$ the transition amplitudes for a radially propagating $\alpha$ neutrino to be detected as an electron neutrino $\nu_e$, a muon neutrino $\nu_\mu$ or a tau neutrino $\nu_\tau$ at a given radial coordinate $r$:
\begin{equation}\label{ColumnVector}
\chi(r)=
\begin{pmatrix}
\braket{\nu_e|\nu_\alpha(r)}\\
\braket{\nu_\mu|\nu_\alpha(r)}\\
\braket{\nu_\tau|\nu_\alpha(r)}
\end{pmatrix}.
\end{equation}
Having already found the contribution of gravity to each of the transition amplitudes in Eq.\,(\ref{VacuumTransitionAmplitudes}) of Sec.\,\ref{Sec:Vacuum}, we only need to add in here the effect of matter as it arises from the effective potential (\ref{ConfVCC}) in the manner done in Ref.\,\cite{Heuristic}. Thus, the transition amplitudes $\chi(r)$ of conformally coupled neutrinos at a point $r$ are related to the transition amplitudes $\chi(r_0)$ at point $r_0$ by
\begin{widetext}
\begin{align}\label{TransitionVectorAtB}
\chi(r)&=\exp\left(-i\,U\left[\frac{m_*^2}{2E_0}\mathds{1}+\frac{\mathbb{M}^2}{2E_0}\right]U^\dagger\int_{r_0}^{r}\Omega(\phi)\sqrt{\mathcal{A}(r)\mathcal{B}(r)}\,{\rm d}r-i\int_{0}^{t}\tilde{V}_{\rm CC}(r,\phi)\,{\rm d}t\right)\chi(r_0)\nonumber\\
&=\exp\left(-i\,U\frac{\mathbb{M}^2}{2E_0}U^\dagger\int_{r_0}^{r}\Omega(\phi)\sqrt{\mathcal{A}(r)\mathcal{B}(r)}\,{\rm d}r-i\int_{r_0}^{r}\tilde{V}_{\rm CC}(r,\phi)\sqrt{\frac{\mathcal{B}(r)}{\mathcal{A}(r)}}{\rm d}r\right)\chi(r_0)\nonumber\\
&=\exp\left[-\frac{i}{2E_0}\left(\mathbb{M}^2_f\int_{r_0}^{r}\Omega(\phi)\sqrt{\mathcal{A}(r)\mathcal{B}(r)}\,{\rm d}r+\mathbb{A}_{\rm CC}\int_{r_0}^{r}\Omega(\phi)n_e(r)\sqrt{\frac{\mathcal{B}(r)}{\mathcal{A}(r)}}{\rm d}r\right)\right]\chi(r_0).
\end{align}
\end{widetext}
In the first step, we inserted the accumulated quantum phases (\ref{trPhase}) of each mass eigenstate and expressed them in the flavor basis thanks to the mixing matrix $U$. The mass matrix $\mathbb{M}$ is the diagonal matrix, ${\rm diag}\,(m_1^2,m_2^2,m_3^2)$, written in the mass-eigenstate basis, and $\mathds{1}$ is the $3\times3$ unit matrix. In the second step, we factored out and discarded the phase factor $\exp(-im_*^2/2E_0)$, which is common to all flavors, and we used Eq.\,(\ref{dr/dt}) to convert the integral over time into an integral over space keeping only the leading order. In the last step, we introduced the notation $\mathbb{M}_f$ for the mass matrix in the flavor basis and we introduced, for convenience, the matrix notation
\begin{equation}\label{AMatrix}
    \mathbb{A}_{\rm CC}=
    \begin{pmatrix}
    v & 0 & 0\\
    0 & 0 & 0\\
    0 & 0 & 0
    \end{pmatrix},
\end{equation}
where $v=2\sqrt{2}E_0G_F$. The result (\ref{TransitionVectorAtB}) suggests that the dynamics of the transition amplitudes has the following Schr\"odinger-like equation structure in matrix form:
\begin{align}\label{SchrodingerLike}
i\frac{\rm d}{{\rm d}r}\chi(r)&= \frac{\Omega(\phi)}{2E_0}\Bigg[\mathbb{M}_f^2\sqrt{\mathcal{A}(r)\mathcal{B}(r)}+\mathbb{A}_{\rm CC}\,n_e(r)\sqrt{\frac{\mathcal{B}(r)}{\mathcal{A}(r)}}\Bigg]\Psi(r)\nonumber\\
&\equiv\mathcal{H}_F\,\chi(r).
\end{align}
We have introduced here the notation $\mathcal{H}_F$ for the effective Hamiltonian in the flavor basis. We shall apply this result to the simple and more instructive case of two-flavor neutrinos in the next section.

\section{Two-flavor neutrinos case}\label{Sec:Twoflavor}
For the sake of clarity and simplicity, we apply here the previous results to the case of two flavors and two mass eigenstates so that $j,k\in\{1,2\}$. Furthermore, we consider here only the transitions between the electron neutrino $\nu_e$ and the muon neutrino $\nu_\mu$, so that $\alpha,\beta\in\{e,\mu\}$. The $\nu_e-\nu_\tau$ transition probability can be computed in exactly the same manner. These transitions are also the most relevant ones in view of their applications to solar neutrinos. 

In the two-flavor neutrinos case, the mixing matrix $U^*_{\alpha j}$ has the following form in terms of the mixing angle $\theta$ \cite{NBook}:
\begin{equation}\label{UMatrix}
    U^*_{\alpha j}=
    \begin{pmatrix}
    \cos\theta & \sin\theta \\
    -\sin\theta & \cos\theta
    \end{pmatrix}.
\end{equation}
For the electron and muon neutrinos, the column vector (\ref{ColumnVector}) for the transition amplitudes reads
\begin{equation}\label{2FlavorColumnVector}
\chi(r)=
\begin{pmatrix}
\braket{\nu_e|\nu_e(r)}\\
\braket{\nu_\mu|\nu_e(r)}
\end{pmatrix}.
\end{equation}
We also denote here, for convenience, the squared-mass difference by $\Delta m^2\equiv m_2^2-m_1^2$. Then, by factoring out and discarding the diagonal matrix $\frac{\Omega(\phi)}{2E_0}\sqrt{\frac{\mathcal{B}(r)}{\mathcal{A}(r)}}\left[\left(\frac{1}{2}\Delta m+m_1^2\right)\mathcal{A}(r)-\frac{1}{2}vn_e(r)\right]\mathds{1}$, which gives rise to an irrelevant phase factor common to both flavors, the effective Hamiltonian matrix in Eq.\,(\ref{SchrodingerLike}) takes the following form
\begin{widetext}
\begin{equation}\label{2FlavorHamiltonian}
\mathcal{H}_F=\frac{\Omega(\phi)}{4E_0}\sqrt{\dfrac{\mathcal{B}(r)}{\mathcal{A}(r)}}
\begin{pmatrix}
-\mathcal{A}(r)\Delta m^2\cos2\theta+v n_e(r)  & \mathcal{A}(r)\Delta m^2\sin2\theta\\
\mathcal{A}(r)\Delta m^2\sin2\theta & \mathcal{A}(r)\Delta m^2\cos2\theta-v n_e(r)
\end{pmatrix}.
\end{equation}
\end{widetext}
We can diagonalize this matrix by the orthogonal transformation $U_M^\dagger\mathcal{H}_F U_M$, so that the new Hamiltonian matrix takes the form $\mathcal{H}_M=\frac{\Omega(\phi)}{4E_0}{\rm diag}(-\Delta M^2,\Delta M^2)$, where $\Delta M^2\equiv M_2^2-M_1^2$ is the squared-mass difference corresponding to the effective masses $M_1$ and $M_2$ of the mass eigenstates inside matter and under a non-negligible gravitational field. The unitary matrix $U_M$ for achieving this might be written as,
\begin{equation}\label{EffectUM}
    U_M=
    \begin{pmatrix}
    \cos\theta_M & \sin\theta_M \\
    -\sin\theta_M & \cos\theta_M
    \end{pmatrix}.
\end{equation}
The new angle $\theta_M$ would thus represent the effective mixing angle for conformally coupled neutrinos in matter inside a gravitational field. A straightforward calculation then shows that the mixing angle $\theta_M$ in matter for diagonalizing the Hamiltonian (\ref{2FlavorHamiltonian}) is given by
\begin{equation}\label{TangentThetaM}
\tan 2\theta_M=\frac{\mathcal{A}(r)\Delta m^2\sin2\theta}{\mathcal{A}(r)\Delta m^2\cos2\theta-vn_e(r)},
\end{equation}
which, in turn, leads to the following expression for the effective squared-mass difference:
\begin{align}\label{DeltaM}
    \Delta M^2&=\sqrt{\frac{\mathcal{B}(r)}{\mathcal{A}(r)}}\nonumber\\
    &\times\Bigg(\left[\mathcal{A}(r)\Delta m^2\cos2\theta-vn_e(r)\right]^2+\left[\mathcal{A}(r)\Delta m^2\sin2\theta\right]^2\Bigg)^{\frac{1}{2}}.
\end{align}
From the latter expression, we deduce the following resonance condition that makes the effective squared-mass difference $\Delta M^2$ reach its minimum and the effective mixing angle take the value $\theta_M=\frac{\pi}{4}$ for which the mixing is maximal:
\begin{equation}\label{2FlavorElectronDensity}
    n_e(r)=\frac{\mathcal{A}(r)\Delta m^2}{2\sqrt{2}E_0G_F}\cos2\theta.
\end{equation}
This result describes how the conformal coupling modifies the familiar MSW effect inside matter and under a gravitational field. To see the explicit contribution of the conformal coupling, we express this identity in terms of the density $\tilde{n}_e(r)={\rm d}N_e/{\rm d}V$ relative to the metric $g_{\mu\nu}$ as follows:
\begin{equation}\label{2FlavorElectronDensityBis}
    \tilde{n}_e(r)=\frac{\mathcal{A}(r)}{\Omega^3(\phi)}\frac{\Delta m^2}{2\sqrt{2}E_0G_F}\cos2\theta.
\end{equation}
By combining Eqs.\,(\ref{TangentThetaM}) and (\ref{DeltaM}), we find
\begin{align}\label{CosSinThetaM}
    \cos2\theta_M&=\sqrt{\frac{\mathcal{B}(r)}{\mathcal{A}(r)}}\frac{\mathcal{A}(r)\Delta m^2\cos2\theta-vn_e(r)}{\Delta M^2},\nonumber\\
    \sin2\theta_M&=\sqrt{\frac{\mathcal{B}(r)}{\mathcal{A}(r)}}\frac{\mathcal{A}(r)\Delta m^2\sin2\theta}{\Delta M^2}.
\end{align}
Using these two expressions, the effective Hamiltonian (\ref{2FlavorHamiltonian}) takes the following simpler form in terms of the effective squared-mass $\Delta M^2$ and effective mixing 
angle $\theta_M$:
\begin{equation}\label{Simpler2FlavorHamiltonian}
\mathcal{H}_F=\frac{\Omega(\phi)}{4E_0}
\begin{pmatrix}
-\Delta M^2\cos2\theta_M  & \Delta M^2\sin2\theta_M\\
\Delta M^2\sin2\theta_M & \Delta M^2\cos2\theta_M
\end{pmatrix}.
\end{equation}
Next, using the unitary matrix (\ref{EffectUM}),  we can express the transition-amplitudes vector (\ref{2FlavorColumnVector}) in the effective mass eigenstates basis in matter, $(\ket{\nu_{_{M1}}},\ket{\nu_{_{M2}}})$, by the transformation
\begin{equation}\label{EffectiveEigenstates}
    \Theta(r)\equiv
    \begin{pmatrix}
    \braket{\nu_{_{M1}}|\nu_e(r)}\\
    \braket{\nu_{_{M2}}|\nu_e(r)}
    \end{pmatrix}=U_M^\dagger\,\chi(r),
\end{equation}
so that the effective Hamiltonian (\ref{Simpler2FlavorHamiltonian}) leads to the following Schr\"odinger-like dynamical equation for the column vector $\Theta(r)$:
\begin{equation}\label{EffectiveSchrodinger-Like}
i\frac{\rm d}{{\rm d}r}\Theta(r)=
\begin{pmatrix}
-\Omega(\phi)\Delta M^2/4E_0  & -i{\rm d}\theta_M/{\rm d}r\\
i{\rm d}\theta_M/{\rm d}r & \Omega(\phi)\Delta M^2/4E_0
\end{pmatrix}\Theta(r).
\end{equation}

The off-diagonal terms proportional to ${\rm d}\theta_M/{\rm d}r$ in this matrix equation are responsible for generating transitions between the effective mass eigenstates $\ket{\nu_{_{M1}}}$ and $\ket{\nu_{_{M2}}}$. Now, in contrast to the flat-spacetime case, the derivative ${\rm d}\theta_M/{\rm d}r$ does not vanish even for a constant electrons number density $n_e(r)$. Indeed, the presence of the time-component $\mathcal{A}(r)$ of the metric in expression (\ref{TangentThetaM}) would still prevent the effective mixing angle $\theta_M$ from being uniform. In fact, the condition for having ${\rm d}\theta_M/{\rm d}r=0$ translates here into the following condition relating the variation of the number density $n_e(r)$ and the component $\mathcal{A}(r)$ of the metric:
\begin{equation}\label{ThetaMConstant}
    \frac{1}{n_e(r)}\frac{{\rm d}n_e(r)}{{\rm d}r}=\frac{1}{\mathcal{A}(r)}\frac{{\rm d}\mathcal{A}(r)}{{\rm d}r}.
\end{equation}
Using the relation $\tilde{n}_e(r)=\Omega^3(\phi)n_e(r)$, we may express now this condition in terms of the number density $\tilde{n}_e(r)={\rm d}N_e/{\rm d}V$ that involves the actual metric of spacetime $g_{\mu\nu}$ instead of $\tilde{g}_{\mu\nu}$:
\begin{equation}\label{ThetaMConstantBis}
    \frac{1}{\tilde{n}_e(r)}\frac{{\rm d}\tilde{n}_e(r)}{{\rm d}r}=\frac{1}{\mathcal{A}(r)}\frac{{\rm d}\mathcal{A}(r)}{{\rm d}r}+\frac{3}{\Omega(\phi)}\frac{{\rm d}\Omega(\phi)}{{\rm d}r}.
\end{equation}
The term on the left-hand side of this equation arises from the usual variation of the density of the electrons making the medium, regardless whether those electrons are conformally coupled or not. That term depends only on the nature of the medium. The second term on the right-hand side, however, depends only on the conformal coupling, and hence only on the variation of the scalar field $\phi$ inside the medium. When the specific condition (\ref{ThetaMConstantBis}) is satisfied, the matrix in the dynamical equation (\ref{EffectiveSchrodinger-Like}) reduces to a diagonal matrix leading to a decoupling of the effective neutrino mass eigenstates, which results then in the following transition probability in matter
\begin{equation}\label{2FlavorMatterPWProba}
    \mathcal{P}^{M}_{\nu_e\rightarrow\nu_\mu}(r)= \sin^22\theta_M\sin^2\left(\int_{r_0}^r\frac{\Omega(\phi)\Delta M^2}{4E_0}{\rm d}r\right).
\end{equation}
The oscillation length in this case is therefore given by
\begin{equation}\label{MatterLength}
  L_{\rm osc}^M=4\pi E_0\left(\int_{r_0}^r\Omega(\phi)\Delta M^2{\rm d}r\right)^{-1}.  
\end{equation}
These expressions are similar in form to those in vacuum, except that the squared-mass difference $\Delta m^2$ is replaced here by the effective $\phi$-dependent squared-mass difference $\Delta M^2$. Note also that our transition probability (\ref{2FlavorMatterPWProba}) never exceeds unity, in contrast to what is found in Refs.\,\cite{PRD2021a,Arxiv2022}.

On the other hand, when condition (\ref{ThetaMConstantBis}) is not satisfied the off-diagonal terms in Eq.\,(\ref{EffectiveSchrodinger-Like}) give rise to transitions between the effective mass eigenstates $\ket{\nu_{M_1}}$ and $\ket{\nu_{M_2}}$. These transitions remain negligible only when the adiabaticity parameter $\gamma$, defined by
\begin{equation}\label{Adiabaticity}
\gamma\equiv\frac{\Omega(\phi)\Delta M^2}{4E_0\,|{\rm d}\theta_M/{\rm d}r|},    
\end{equation}
is much larger than unity. Using Eqs.\,(\ref{TangentThetaM}) and (\ref{DeltaM}), this adiabaticity parameter takes the following more explicit form
\begin{equation}\label{ExplicitAdiabatic}
\gamma=\frac{\Omega(\phi)(\Delta M^2)^2}{4\sqrt{2}E_0G_Fn_e(r)\sin2\theta_M}\sqrt{\frac{\mathcal{A}(r)}{\mathcal{B}(r)}}\left[\frac{n'_e(r)}{n_e(r)}-\frac{\mathcal{A}'(r)}{\mathcal{A}(r)}\right]^{-1}.
\end{equation}
Here, the primes stand for a derivative with respect to the radial coordinate $r$. This expression shows that when the variation of the gravitational field with position is taken into account, the adiabaticity parameter could reach a very large value even when the derivative of the electron number density does not vanish anywhere, which is in contrast to what happens in the Minkowski spacetime.

\subsection{Application to the chameleon and symmetron models}
We apply here our formulas (\ref{2FlavorElectronDensity}), (\ref{ThetaMConstantBis}) and (\ref{ExplicitAdiabatic}) to the case of the chameleon and symmetron models. First of all, note that according to Eq.\,(\ref{2FlavorElectronDensity}) an oscillation resonance cannot occur if the number density of electrons $n_e(r)$ is uniform inside the compact object through which the neutrinos are traveling. This is in contrast to the case where gravity is neglected, and it is due to the presence of the radially varying metric component $\mathcal{A}(r)$ on the right-hand side of Eq.\,(\ref{2FlavorElectronDensity}). This being the case both with and without a conformal coupling, for in the presence of the conformal coupling Eq\,(\ref{2FlavorElectronDensity}) reads in terms of the number density $\tilde{n}_e(r)$ in the chameleon and symmetron models, respectively, as follows:
\begin{align}\label{Chameleon2FlavorElectronDwnsity}
     \tilde{n}_e(r)&=\frac{\mathcal{A}(r)\Delta m^2}{2\sqrt{2}E_0G_F}\exp\left[-3\beta\phi(r)\right]\cos2\theta,\nonumber\\
     \tilde{n}_e(r)&=\frac{\mathcal{A}(r)\Delta m^2}{2\sqrt{2}E_0G_F}\left[1+\beta\phi^2(r)\right]^{-3}\cos2\theta.
\end{align}
Setting $\tilde{n}_e(r)$ equal to a constant in these identities would indeed entail one extra relation between the metric component and the scalar field which is not part of the equation of motion for the scalar field in either models (see e.g., Ref.\,\cite{ChameleonIntroduction}). These two identities show then that the required number density of electrons for a resonance to occur is dictated by the scalar field's profile as well as the interior gravitational field inside the compact object.

Similarly, Eq.\,(\ref{ThetaMConstant}) takes the following form for the chameleon and symmetron models, respectively:
\begin{align}\label{ChameleonSymmetronThetaMConstantBis}
    \frac{1}{\tilde{n}_e(r)}\frac{{\rm d}\tilde{n}_e(r)}{{\rm d}r}&=\frac{1}{\mathcal{A}(r)}\frac{{\rm d}\mathcal{A}(r)}{{\rm d}r}+3\beta\frac{{\rm d}\phi(r)}{{\rm d}r},\nonumber\\
    \frac{1}{\tilde{n}_e(r)}\frac{{\rm d}\tilde{n}_e(r)}{{\rm d}r}&=\frac{1}{\mathcal{A}(r)}\frac{{\rm d}\mathcal{A}(r)}{{\rm d}r}+\frac{6\beta\phi(r)}{1+\beta\phi^2(r)}\frac{{\rm d}\phi(r)}{{\rm d}r}.
\end{align}
These two differential equations relate in each model the radial variation of the electron number density to the radial variation of the metric component and the scalar field inside the compact object for a decoupling between the effective mass eigenstates $\ket{\nu_{M_1}}$ and $\ket{\nu_{M_2}}$ to occur.  
According to these two conditions, we therefore conclude that in the presence of a uniform number density $\tilde{n}_e(r)$ no decoupling between the effective mass eigenstates would result. Furthermore, because of the presence of the derivative of the metric component $\mathcal{A}(r)$ on the right-hand side of each equation, this conclusion remains valid even in the absence of any conformal coupling of the neutrinos.

Finally, the adiabaticity parameter $\gamma$, which measures how decoupled the mass eigenstates become, takes the following explicit form in the chameleon model and symmetron model, respectively:
\begin{align}\label{ChameleonSymmetronExplicitAdiabatic}
\gamma&=\frac{e^{4\beta\phi(r)}(\Delta M^2)^2}{4\sqrt{2}E_0G_F\tilde{n}_e(r)\sin2\theta_M}\sqrt{\frac{\mathcal{A}(r)}{\mathcal{B}(r)}}\nonumber\\
&\quad\qquad\qquad\qquad\qquad\qquad\times\left[\frac{\tilde{n}'_e(r)}{\tilde{n}_e(r)}-3\beta\phi'(r)-\frac{\mathcal{A}'(r)}{\mathcal{A}(r)}\right]^{-1},\nonumber\\
\gamma&=\frac{[1+\beta\phi^2(r)]^4(\Delta M^2)^2}{4\sqrt{2}E_0G_F\tilde{n}_e(r)\sin2\theta_M}\sqrt{\frac{\mathcal{A}(r)}{\mathcal{B}(r)}}\nonumber\\
&\qquad\qquad\qquad\qquad\qquad\times\left[\frac{\tilde{n}'_e(r)}{\tilde{n}_e(r)}-\frac{6\beta\phi(r)\phi'(r)}{1+\beta\phi^2(r)}-\frac{\mathcal{A}'(r)}{\mathcal{A}(r)}\right]^{-1}\!\!\!.
\end{align}
The presence of the derivative of the term containing the scalar field and its derivative in the denominator in these expressions shows that in both models the adiabaticity parameter could reach very larger values even in the absence of the gravitational field, i.e., even in Minkowski spacetime.
\section{The non-universal conformal coupling case}\label{Sec:AppB}
We shall examine now how our previous results get modified when the conformal coupling of neutrinos is allowed to be non-universal. A conformal coupling is said to be non-universal when the scalar field $\phi$ and/or the conformal factor $\Omega(\phi)$ have different profiles/forms for different mass eigenstates of the neutrinos. We assume both possibilities here by simply adding a subscript $j$ and denoting by $\Omega_j(\phi)$ the conformal factor that couples the $j$-th mass eigenstate. 

Including a non-universal conformal coupling in the study of flavor oscillations in matter is much more involved than the corresponding generalization to neutrinos propagating in vacuum conducted in Ref.\,\cite{Submitted}. The reason being that one has to start by generalizing the tree-level Lagrangian (\ref{ConfLagrangian}) by introducing a different conformal coupling $\Omega(\phi)$ for each of the fields involved in that Lagrangian. However, given that the Lagrangian (\ref{ConfLagrangian}) involves five different field masses, we need to introduce five different conformal factors: one for the electron field $\psi_e(x)$, three for the neutrino mass eigenstates $\psi_j(x)$ making the flavor $\psi_{\nu_e}(x)$, and one for the $W$ bosons' field $W_\mu(x)$. Therefore, it is not possible anymore to just replace the spacetime metric $g_{\mu\nu}$ in that matter Lagrangian by the conformal metric $\tilde{g}_{\mu\nu}$, for no such common metric to all the involved fields exists. 

The way around would be to keep the original metric $g_{\mu\nu}$ in that Lagrangian and replace each of the five fields by their conformal counterparts. This procedure will indeed be implemented without any particular difficulty. However, we should emphasize here that such a procedure departs from the standard recipe that consists in first assigning to the fields a common conformal metric and then switching back to the original metric by replacing all of the fields by their conformal counterparts. Bearing in mind this important remark, we proceed now to derive the new version of the effective Hamiltonian inside matter in Eq.\,(\ref{SchrodingerLike}) that corresponds to the non-universal conformal coupling scenario.

We start by finding a new version for Eq.\,(\ref{TransitionVectorAtB}). Taking into account the five different conformal couplings involved, the transition amplitudes $\chi(r)$ at a point $r$ would be related to the transition amplitudes $\chi(r_0)$ at point $r_0$ by
\begin{widetext}
\begin{align}\label{Sec6:TransitionVectorAtB}
\chi(r)&=\exp\left(-i\int_{r_0}^{r}U\left[\frac{\Omega_*(\phi)m_*^2}{2E_0}\mathds{1}+\frac{\mathbb{M}^2(\phi)}{2E_0}\right]U^\dagger\sqrt{\mathcal{A}(r)\mathcal{B}(r)}\,{\rm d}r-i\int_{0}^{t}\tilde{V}_{\rm CC}(r,\phi)\,{\rm d}t\right)\chi(r_0)\nonumber\\
&=\exp\left[-\frac{i}{2E_0}\left(\int_{r_0}^{r}\mathbb{M}^2_f(\phi)\sqrt{\mathcal{A}(r)\mathcal{B}(r)}\,{\rm d}r+\mathbb{A}_{\rm CC}\int_{r_0}^{r}\Omega^{-2}_W(\phi)\Omega^3_e(\phi)n_e(r)\sqrt{\frac{\mathcal{B}(r)}{\mathcal{A}(r)}}{\rm d}r\right)\right]\chi(r_0).
\end{align}
\end{widetext}
In the first step, we took care of distinguishing the conformal coupling of each mass eigenstate. Thus, the factor $\Omega_*(\phi)$ corresponds to the most massive of the states and the mass matrix $\mathbb{M}(\phi)$ is now the diagonal matrix ${\rm diag}\,\left[\Omega_1(\phi)m_1^2,\Omega_2(\phi)m_2^2,\Omega_3(\phi)m_3^2\right]$. In the second step, we again factored out and discarded the phase factor $\exp\left[-i\Omega_*(\phi)m_*^2/2E_0\right]$, which is common to all flavors, and we introduced the notation $\mathbb{M}_f(\phi)$ for the new mass matrix $\mathbb{M}(\phi)$ in the flavor basis. The matrix $\mathbb{A}_{\rm CC}$ is still given here by Eq.\,(\ref{AMatrix}). The two additional conformal factors $\Omega_W(\phi)$ and $\Omega_e(\phi)$ in the second integral come from the fact that the effective potential $\tilde{V}_{\rm CC}(x,\phi)$ of Eq.\,(\ref{ConfVCC}) is now given by
\begin{equation}\label{Sec6:ConfVCC}
    \tilde{V}_{\rm CC}(x,\phi)=\sqrt{2}\,\tilde{G}_F\tilde{n}_e(x)=\sqrt{2}\,G_F\,\Omega^{-2}_W(\phi)\Omega^3_e(\phi)n_e(x).
\end{equation}
The conformal factor $\Omega_W^{-2}(\phi)$ in the second step arises from the transformation of the $W$ boson's mass, $\tilde{m}_W=\Omega_W(\phi)m_W$, which implies that Eq.\,(\ref{ConfEffLagrangian}) yields $\tilde{G}_F=\frac{\sqrt{2}}{8}g_W^2/\tilde{m}_W^2=\Omega^{-2}_W(\phi)G_F$. The conformal factor $\Omega_e^3(\phi)$ arises from the fact that the electron number density transforms as $\tilde{n}_e(x)=\Omega^3_e(\phi)n_e(x)$. This transformation is a consequence of assigning the conformal factor $\Omega_e(\phi)$ to the infinitesimal spatial volume containing the number $N_e$ of electrons so that ${\rm d}N_e/{\rm d}V=\Omega^3_e(\phi){\rm d}N_e/{\rm d}\tilde{V}$.   

Therefore, the new version of the effective Hamiltonian inside matter reads,
\begin{multline}\label{Sec6:SchrodingerLike}
\mathcal{H}_F= \frac{1}{2E_0}\Bigg[\mathbb{M}_f^2(\phi)\sqrt{\mathcal{A}(r)\mathcal{B}(r)}\\
+\mathbb{A}_{\rm CC}\,\Omega_W^{-2}(\phi)\Omega_e^3(\phi)n_e(r)\sqrt{\frac{\mathcal{B}(r)}{\mathcal{A}(r)}}\Bigg].
\end{multline}
For simplicity, we shall henceforth restrict again our analysis in this section to the case of two-flavor neutrinos. For that purpose, we introduce the $\phi$-dependent squared-mass difference  $\Delta m^2(\phi)\equiv\Omega_2(\phi)m_2^2-\Omega_1(\phi)m_1^2$. Then, we factor out and discard from the right-hand side of Eq.\,(\ref{Sec6:SchrodingerLike}) the diagonal matrix $\frac{1}{2E_0}\!\!\sqrt{\frac{\mathcal{B}(r)}{\mathcal{A}(r)}}\left[\left(\frac{1}{2}\Delta m(\phi)+\Omega_1(\phi)m_1^2\right)\mathcal{A}(r)\!-\!\frac{1}{2}\Omega^{-2}_W(\phi)\Omega^3_e(\phi)vn_e(r)\right]\!\mathds{1}$, which gives rise to an irrelevant phase factor common to both flavors. Therefore, for two-flavor neutrinos the effective Hamiltonian (\ref{Sec6:SchrodingerLike}) reads,
\begin{widetext}
\begin{equation}\label{Sec6:2FlavorHamiltonian}
\mathcal{H}_F=\frac{1}{4E_0}\sqrt{\dfrac{\mathcal{B}(r)}{\mathcal{A}(r)}}
\begin{pmatrix}
-\mathcal{A}(r)\Delta m^2(\phi)\cos2\theta+v\, \dfrac{\Omega^3_e(\phi)}{\Omega^{2}_W(\phi)}n_e(r)  & \mathcal{A}(r)\Delta m^2(\phi)\sin2\theta\\
\mathcal{A}(r)\Delta m^2(\phi)\sin2\theta & \mathcal{A}(r)\Delta m^2(\phi)\cos2\theta-v\,\dfrac{\Omega^3_e(\phi)}{\Omega^{2}_W(\phi)} n_e(r)
\end{pmatrix}.
\end{equation}
\end{widetext}
Following the same steps that allowed us to diagonalize the matrix (\ref{2FlavorHamiltonian}) in Sec.\,\ref{Sec:Twoflavor}, we learn that the mixing angle $\theta_M$ in the unitary matrix (\ref{EffectUM}) is given here by
\begin{equation}\label{Sec6:TangentThetaM}
\tan 2\theta_M=\frac{\mathcal{A}(r)\Delta m^2(\phi)\sin2\theta}{\mathcal{A}(r)\Delta m^2(\phi)\cos2\theta-v\,\Omega^{3}_e(\phi)n_e(r)/\Omega^{2}_W(\phi)},
\end{equation}
which, in turn, leads to the following expression for the effective squared-mass difference:
\begin{multline}\label{Sec6:DeltaM}
    \Delta M^2=\sqrt{\frac{\mathcal{B}(r)}{\mathcal{A}(r)}}\Bigg(\left[\mathcal{A}(r)\Delta m^2(\phi)\cos2\theta-v\frac{\Omega^3_e(\phi)}{\Omega^2_W(\phi)}n_e(r)\right]^2\\
    +\left[\mathcal{A}(r)\Delta m^2(\phi)\sin2\theta\right]^2\Bigg)^{\frac{1}{2}}.
\end{multline}
From the latter expression, we deduce the following resonance condition for the case of a non-universal conformal coupling of matter:
\begin{equation}
    n_e(r)=\frac{\mathcal{A}(r)\Delta m^2(\phi)}{2\sqrt{2}E_0G_F}\frac{\Omega^2_W(\phi)}{\Omega^3_e(\phi)}\cos2\theta.
\end{equation}
We see that the non-universality translates into a condition that makes the number density of electrons depend differently on each of the five conformal couplings.

By combining Eqs.\,(\ref{Sec6:TangentThetaM}) and (\ref{Sec6:DeltaM}), we find
\begin{align}\label{Sec6:CosSinThetaM}
    \cos2\theta_M&=\sqrt{\frac{\mathcal{B}(r)}{\mathcal{A}(r)}}\frac{\mathcal{A}(r)\Delta m^2(\phi)\cos2\theta-v\Omega^3_e(\phi)n_e(r)/\Omega^2_W(\phi)}{\Delta M^2},\nonumber\\ \sin2\theta_M&=\sqrt{\frac{\mathcal{B}(r)}{\mathcal{A}(r)}}\frac{\mathcal{A}(r)\Delta m^2(\phi)\sin2\theta}{\Delta M^2}.
\end{align}
Using these two expressions, the effective Hamiltonian (\ref{Sec6:2FlavorHamiltonian}) takes the following form
\begin{equation}\label{Sec6:Simpler2FlavorHamiltonian}
\mathcal{H}_F=\frac{1}{4E_0}
\begin{pmatrix}
-\Delta M^2\cos2\theta_M  & \Delta M^2\sin2\theta_M\\
\Delta M^2\sin2\theta_M & \Delta M^2\cos2\theta_M
\end{pmatrix},
\end{equation}
which leads to the following Schr\"odinger-like dynamical equation for the column vector (\ref{EffectiveEigenstates}) in the effective mass eigenstates basis:
\begin{equation}\label{Sec6:EffectiveSchrodinger-Like}
i\frac{\rm d}{{\rm d}r}\Theta(r)=
\begin{pmatrix}
-\Delta M^2/4E_0  & -i{\rm d}\theta_M/{\rm d}r\\
i{\rm d}\theta_M/{\rm d}r & \Delta M^2/4E_0
\end{pmatrix}\Theta(r).
\end{equation}
Using Eq.\,(\ref{Sec6:TangentThetaM}),
the condition for having here ${\rm d}\theta_M/{\rm d}r=0$ that would prevent transitions between the effective mass eigenstates reads
\begin{equation}\label{Sec6:ThetaMConstant}
    \frac{\mathcal{A}'(r)}{\mathcal{A}(r)}=\frac{n'_e(r)}{n_e(r)}+\frac{3\Omega'_e(\phi)}{\Omega_e(\phi)}+\frac{\Omega'_1(\phi)}{\Omega_1(\phi)}-\frac{\Omega'_2(\phi)}{\Omega_2(\phi)}-\frac{2\Omega'_W(\phi)}{\Omega_W(\phi)}.
\end{equation}
Recalling that $\tilde{n}_e(r)={\rm d}N_e/{\rm d}V$ and that $\tilde{n}_e(r)=\Omega^3_e(\phi)n_e(r)$, this condition can also be written in terms of $\tilde{n}_e(r)$ as,
\begin{equation}\label{Sec6:ThetaMConstantBis}
    \frac{\mathcal{A}'(r)}{\mathcal{A}(r)}=\frac{\tilde{n}'_e(r)}{\tilde{n}_e(r)}+\frac{\Omega'_1(\phi)}{\Omega_1(\phi)}-\frac{\Omega'_2(\phi)}{\Omega_2(\phi)}-\frac{2\Omega'_W(\phi)}{\Omega_W(\phi)}.
\end{equation}
These two differential equations neatly display a difference in the weights associated to the contribution of each of the five conformal couplings. When these differential equations are satisfied, the matrix in the dynamical equation (\ref{Sec6:EffectiveSchrodinger-Like}) reduces to a diagonal matrix leading to a decoupling of the effective neutrino mass eigenstates, which results then in the following transition probability in matter:
\begin{equation}\label{NonUni2FlavorMatterPWProba}
    \mathcal{P}^{M}_{\nu_e\rightarrow\nu_\mu}(r)= \sin^22\theta_M\sin^2\left(\int_{r_0}^r\frac{\Delta M^2}{4E_0}{\rm d}r\right).
\end{equation}
The oscillation length in this case is therefore given by
\begin{equation}\label{NonUniMatterLength}
  L_{\rm osc}^M=4\pi E_0\left(\int_{r_0}^r\Delta M^2{\rm d}r\right)^{-1}.  
\end{equation}
On the other hand, if condition (\ref{Sec6:ThetaMConstantBis}) is not satisfied, the off-diagonal terms in Eq.\,(\ref{Sec6:EffectiveSchrodinger-Like}) give rise to transitions between the effective mass eigenstates $\ket{\nu_{M_1}}$ and $\ket{\nu_{M_2}}$. These transitions remain negligible only when the adiabaticity parameter $\gamma$, given here by
\begin{equation}\label{Adiabaticity}
\gamma\equiv\frac{\Delta M^2}{4E_0\,|{\rm d}\theta_M/{\rm d}r|},    
\end{equation}
is much larger than unity. Using Eqs.\,(\ref{Sec6:TangentThetaM}) and (\ref{Sec6:DeltaM}), this adiabaticity parameter takes the following more explicit form:
\begin{multline}\label{sec6:ExplicitAdiabatic}
\gamma=\frac{(\Delta M^2)^2\Omega^2_e(\phi)/\Omega^3_W(\phi)}{4\sqrt{2}E_0G_Fn_e(r)\sin2\theta_M}\sqrt{\frac{\mathcal{A}(r)}{\mathcal{B}(r)}}\\
\times\left[\frac{n'_e(r)}{n_e(r)}+\frac{3\Omega'_e(\phi)}{\Omega_e(\phi)}+\frac{\Omega'_1(\phi)}{\Omega_1(\phi)}-\frac{\Omega'_2(\phi)}{\Omega_2(\phi)}-\frac{2\Omega'_W(\phi)}{\Omega_W(\phi)}-\frac{\mathcal{A}'(r)}{\mathcal{A}(r)}\right]^{-1}.
\end{multline}
This expression displays different contribution weights for the four different coupling factors. Therefore, it is in principle possible to distinguish between different non-universal coupling patterns even within the same screening model. All our expressions in this section reduce, of course, to the universal coupling case when all the conformal factors become identical. 
\section{Conclusion and discussion}\label{Sec:Conclusion}
We studied the MSW effect on the flavor oscillations of conformally coupled neutrinos inside matter in the presence of a non-negligible gravitational field. We derived the modified resonance condition, and we extracted the modified adiabaticity parameter of the MSW effect. We have restricted our study to the tree-level interaction because only the low-energy effective Lagrangian is required to extract the effective Fermi constant under the conformal coupling. However, an extended study involving perturbation theory under the conformal coupling is left for future work.

Although screening models make it possible to violate the equivalence principle, even when their scalar field universally couples to matter, we have nevertheless explored in this work the additional possibility of having a non-universal conformal coupling in those models. A non-universal coupling offers indeed an additional way for violating the equivalence principle. We derived the general formulas for the transition probability for such a case both in matter and in vacuum. Furthermore, we stressed out that the transition probabilities derived here never exceed unity ---\,in contrast to what is found in some recent works from the literature\,--- because a conformal transformation when properly implemented leads to the conservation of quantum mechanics' unitarity. 

Finally, we have restricted our study in this paper only to the flavor oscillations of neutrinos. The effect of conformal couplings on the interesting spin oscillations of neutrinos will be discussed in a forthcoming paper.

\section*{Acknowledgments}
This work was supported by the Natural Sciences and Engineering Research Council of Canada (NSERC) Discovery Grant No. RGPIN-2017-05388; and by the Fonds de Recherche du Québec - Nature et Technologies (FRQNT). PS acknowledges support from Bishop's University through the Graduate Entrance Scholarship award and through the Research Assistantship award.\\\\

\appendix
\section{Derivation of Eq.\,(\ref{ConfVCC})}\label{Sec:AppA}
We derive in this appendix the expression (\ref{ConfVCC}) of the effective potential. For that purpose, we start by using the Fierz identity to rearrange the fermion fields in the effective Lagrangian (\ref{ConfEffLagrangian}) as follows:
\begin{multline}\label{AppA:ConfEffLagrangian}
    \mathcal{\tilde{L}}_{\rm CC}^{\rm eff}=-\frac{\tilde{G}_F}{\sqrt{2}}\sqrt{-g}\left[\bar{\tilde{\psi}}_{\nu_ e}(x)\gamma_\mu\left(1-\gamma^5\right)\tilde{\psi}_{\nu_e}(x)\right]\\
    \times\left[\bar{\tilde{\psi}}_e(x)\gamma^\mu\left(1-\gamma^5\right)\tilde{\psi}_e(x)\right].
\end{multline}
Next, we average this effective Lagrangian density over the electron background in the rest frame of the medium by making use of a statistical distribution function $f(\tilde{E}_{\tilde{\bf p}_e},T)$ of the electrons energy $\tilde{E}_{\tilde{\bf p}_e}$ corresponding to a momentum ${\tilde{\bf p}_e}$ at the temperature $T$ in an unpolarized medium. This distribution function is normalized by $(2\pi)^{-3}\int{\rm d}^3\tilde{\bf p}_e f(\tilde{E}_{\tilde{\bf p}_e},T)=\tilde{n}_e(x)$, where $\tilde{n}_e(x)$ is the number density of the conformally coupled electrons of the medium. Thus we have,
\begin{multline}\label{AppA:AverageConfEffLagrangian}
    \overline{\mathcal{\tilde{L}}_{\rm CC}^{\rm eff}}=-\frac{\tilde{G}_F}{\sqrt{2}}\sqrt{-g}\left[\bar{\tilde{\psi}}_{\nu_ e}(x)\gamma_\mu\left(1-\gamma^5\right)\tilde{\psi}_{\nu_e}(x)\right]\times\\
    \frac{1}{2}\sum_s\int\frac{{\rm d}^3\tilde{\bf p}_e}{(2\pi)^3}\,f(\tilde{E}_{\tilde{\bf p}_e},T)\bra{\tilde{\bf p}_e,s}\bar{\tilde{\psi}}_e(x)\gamma^\mu\left(1-\gamma^5\right)\tilde{\psi}_e(x)\ket{\tilde{\bf p}_e,s}.
\end{multline}
We have inserted the one-electron state $\ket{\tilde{p}_e,s}$ of a conformally coupled electron of momentum $\tilde{p}_e$ and of spin $s$, and then we averaged over the two possible values of the latter. If we define the one-particle state $\ket{\tilde{\bf p}_e,s}$ by
\begin{equation}\label{One-particle}
\ket{\tilde{\bf p}_e,s}=a_{\tilde{\bf p}_e}^{s\dagger}\ket{0},
\end{equation}
and if we expand the electron field $\tilde{\psi}_e(x)$ into the following complete set of modes,
\begin{equation}\label{FieldExpansion}
\tilde{\psi}_e(x)=\!\!\int\frac{{\rm d}^3\tilde{\bf q}_e}{(2\pi)^3}\sum_{r}\!\frac{1}{\sqrt{2\tilde{E}_{\tilde{\bf p}_e}}}\left(a_{\tilde{\bf q}_e}^ru^r(\tilde{q}_e)e^{-i\tilde{q}_e.x}+b_{\tilde{\bf q}_e}^{r\dagger}v^r(\tilde{q}_e)e^{i\tilde{q}_e.x}\right),   
\end{equation}
then the anticommutation relations $\{a^r_{\tilde{\bf p}_e},a^{s\dagger}_{\tilde{\bf q}_e}\}=\{b^r_{\tilde{\bf p}_e},b^{s\dagger}_{\tilde{\bf q}_e}\}=(2\pi)^3\delta^3(\tilde{\bf p}_e-\tilde{\bf q}_e)\,\delta^{rs}$ allow us to write the effective Lagrangian density (\ref{AppA:AverageConfEffLagrangian}) as follows:
\begin{widetext}
\begin{align}\label{AppA:AverageConfEffLagrangianBis}
    \overline{\mathcal{\tilde{L}}_{\rm CC}^{\rm eff}}&=-\frac{\tilde{G}_F}{\sqrt{2}}\sqrt{-g}\left[\bar{\tilde{\psi}}_{\nu_ e}(x)\gamma_\mu\left(1-\gamma^5\right)\tilde{\psi}_{\nu_e}(x)\right]\int\frac{{\rm d}^3\tilde{\bf p}_e}{(2\pi)^3}\frac{f(\tilde{E}_{\tilde{\bf p}_e},T)}{4\tilde{E}_{\tilde{\bf p}_e}}\sum_s\bar{u}^s(\tilde{p}_e)\gamma^\mu\left(1-\gamma^5\right)u^s(\tilde{p}_e)\nonumber\\
    &=-\frac{\tilde{G}_F}{\sqrt{2}}\sqrt{-g}\left[\bar{\tilde{\psi}}_{\nu_ e}(x)\gamma_\mu\left(1-\gamma^5\right)\tilde{\psi}_{\nu_e}(x)\right]\int\frac{{\rm d}^3\tilde{\bf p}_e}{(2\pi)^3}\frac{f(\tilde{E}_{\tilde{\bf p}_e},T)}{4\tilde{E}_{\tilde{\bf p}_e}}{\rm Tr}\left[(-\slashed{\tilde{p}}_e+\tilde{m}_e)\gamma^\mu(1-\gamma^5)\right]\nonumber\\
    &=-\frac{\tilde{G}_F}{\sqrt{2}}\sqrt{-g}\left[\bar{\tilde{\psi}}_{\nu_ e}(x)\gamma_\mu\left(1-\gamma^5\right)\tilde{\psi}_{\nu_e}(x)\right]\int\frac{{\rm d}^3\tilde{\bf p}_e}{(2\pi)^3}\frac{f(\tilde{E}_{\tilde{\bf p}_e},T)}{\tilde{E}_{\tilde{\bf p}_e}}\tilde{p}_e^\mu\nonumber\\
    &=-\sqrt{2}\tilde{G}_F\sqrt{-g}\left[\bar{\tilde{\psi}}_{\nu_{eL}}(x)\gamma^0\tilde{\psi}_{\nu_{eL}}(x)\right]\tilde{n}_e(x).
\end{align}
\end{widetext}
In the last line, we denoted by $\tilde{\psi}_{\nu_{eL}}(x)$ the left-handed electron-neutrino field. From the last line, we easily read off the new effective potential: $\tilde{V}_{\rm CC}(x,\phi)=\sqrt{2}\,\tilde{G}_F\tilde{n}_e(x)$.


\end{document}